# Investigation of Nitrogen Doped Barium Zirconate Using Density Functional Theory

**IFTKHAR AHMAD**

**Department of Physics, Faculty of Basic Sciences and Mathematics, Minhaj University Lahore, Pakistan.**

profiftkharahmad@gmail.com

**Abstract:**

Using density functional theory (DFT), this work explores barium zirconate doped with nitrogen. In addition, we used density functional theory (DFT) to study the $BaZrO_3$'s electrical, optical, and structural properties, and we found that the $BaZrO_3$ has intrinsic vacancy defects by employing the supercell approach. Integrating testability into hardware devices is the goal of testing design, an approach to integrated circuit design that draws on density functional theory. This technology simplifies production testing and implementation. The study strategy incorporates the following methods: LSDA+U, Exchange Correlation Approximations, Hohenberg Kohn Theorem, and Local Spin Density Approximation. We tracked the evolution of barium zirconate's protonic, nitrogen ionic, and electronic conductivities using Density Functional Theory (DFT). We found low electrical conductivity despite the fact that nitrogen doping successfully lowered energy barriers for oxygen and proton ion migration. The conductivity of barium zirconate increased in proportion to the supercell's nitrogen content. Furthermore, we investigated the local arrangements of octahedral sites in Pm3m cubic frameworks and utilized Density Functional Theory to dope $BaZrO_3$. Nitrogen doping caused a lot of changes to oxygen sites, revealing three groups of oxygen atoms with different structural properties.

**Keywords:** Density Functional Theory (DFT), Barium zirconate ($BaZrO_3$), Local Spin Density Approximation (LSDA), Electromagnetic radiation (ER), Doping

**List of Abbreviations:**

Density Functional Theory (DFT), Local Density Approximation (LDA), Generalized Gradient Approximation (GGA), Local Spin Density Approximation (LSDA), Barium Zirconate ($BaZrO_3$), Dynamical Mean Field Theory (DMFT), Around Mean Field (AMF), X-ray Diffraction (XRD), High-Energy-Density (HED), Integrated circuit (IC), Integrated charged a couple (ICC)

## INTRODUCTION

Perovskite materials have great promise for fuel cells, sensors, and batteries. Since the 1980s discovery of strontium cerate ($SrCeO_3$) as a potential proton conductor, scientists have been researching perovskite structured materials for commercially feasible high temperature proton conductors (HTPCs). Doped barium zirconate ($BaZrO_3$) is a good material for high temperature perovskite capacitors (HTPCs) because it is chemically stable and conducts protons well. Its easy proton generation and fast proton movement contribute to its great proton conductivity. In perovskite structures, protons migrate by transferring oxygen ions to their neighbours and rotating around them. Using density functional theory, researchers studied proton movement in bulk $BaZrO_3$ [1]. The bulk undoped $BaZrO_3$ proton transport energy barrier is anticipated to be 0.69 eV. Proton mobility and energy barrier crossing. The rotation barrier was 0.14 eV, and the transfer barrier was 0.23 eV. All ions were allowed to relax during the energy barrier calculation. Barium zirconate is a cubic perovskite because of its larger ionic radius. At room temperature, BZO has a lattice constant of 0.419 nm and a tolerance factor of t = 1.004 [2]. Therefore, it has a very small lattice mismatch with MgO and a similar thermal expansion coefficient. BZO is not ferroelectric, like BTO, because its near perfect ionic radii ratio prevents spontaneous polarization at ambient temperature. Although indirect, the BZO band gap is 5.3 eV, bigger than the BTO band gap of 3.2 eV.

It is found that proton migration energy barriers are 0.18–0.21 eV [3]. However, these recently reported calculated values were substantially lower than experimental values. Multiple explanations have been proposed for the gap between the observed and predicted proton migration energy barriers. They discovered that dopants trapped protons, increasing the energy needed to move them between atoms. The figures were lower, but they were close to experimental energy limits. By thinking fast protons won't use the lowest energy pathways, to set higher energy barriers [4]. DFT calculations with lattice relaxation restrictions yielded results that met experimental limitations. A novel method for finding long-range proton transport channels utilizes color coding and kinetic Monte Carlo. The average limiting barrier was changed to 0.42 eV for a seven-step periodic route at 600 K and to 0.33 eV for an eight-step route. This was done by adding lattice rearrangement barriers. Even though the five- or six-step periodic technique had a barrier closer to the experimental value, it was unclear how to account for variations in periodic route length. When creating samples with varying compositions, verifying sample stoichiometry is critical. This study uses thermogravimetric analysis and X-ray photoelectron spectroscopy. Electromagnetic radiation interactions largely determine a substance's optical characteristics. Because of this, oscillating electric or magnetic fields demand a reaction. The greater coulomb interaction causes most electromagnetic radiation interactions to be electric. Low frequencies excite ionic cores, whereas high frequencies excite electrons. Depending on the radiation frequency, an atom's optical characteristics may reveal its electrons and ionic cores [5]. The visible and ultraviolet spectra are dominated by electron interactions. They have describes the two methods utilised in this research to study electronic transitions and material optical properties. These methods are absorption spectroscopy and variable-angle spectroscopic ellipsometry. Material structure determines several electrical properties. A single structural characterization technique is unnecessary, as structures can be described on different length scales. This alloying enhances the optical characteristics of $BaZrO_3$, making it beneficial in manufacturing and industry. Doped $BaZrO_3$ is a solid perovskite material with good proton conductivity at 400–650 degrees Celsius. This substance is chemically stable. It will enhance perovskites' efficiency and stability. Ceramics made of barium zirconate ($BaZrO_3$) are being thought about for use in protonic fuel cells and hydrogen separation membranes.

Introduction to density functional theory (DFT) recent computational material science research has exploded in fundamental and applied areas. One of the most commonly used ab initio approaches for calculating atom, molecule, crystal, surface, and interaction structures is density functional theory (DFT) [6]. This alternate introduction to DFT uses thermodynamic concepts, particularly switching between independent variables. In DFT, the idea of replacing external potential with density distribution is presented as a simple generalisation of the Legendre transform from chemical potential μ to particle number N [7]. This method introduces the Hohenberg–Kohn energy functional and its theorems, using classical nonuniform fluids as easy examples. Next, the Kohn–Sham equations are obtained from the electronic system energy functional. They look at the exchange-correlation part of this function, including how to get a good approximation of its local density and how to write it exactly in terms of the exchange-correlation hole.

From the literature review, Solids' state physics have physical properties depend on their electrical structure. The Thomas-Fermi model introduced a density functional, but it was abandoned due to its drawbacks. The present KS-DFT uses Walter Kohn and Sham's practical technique to solve the Schrodinger equation [8]. This approach, as well as more exact exchange-correlation energy functionals, has been used to determine the ground-state characteristics of periodic solids, clusters, quantum dots, and molecules. For those interested, the following sections provide an overview of DFT and reference important papers and research. The variational principle and perturbation theory are applied to the KS energy functional. Traditional variational perturbation theorems are also used. Emphasise Green's function and sum over state-based perturbation expansions of wave functions. Use the explicit equations for variational principles for any level of perturbation to obtain a universal formulation for higher order energy derivatives [9]. Observable physical characteristics are listed in order of total energy derivatives. First-order components include the dipole moment, force, and stress tensor. This phenomenon involves elements such as second-order phonon dynamical matrices, piezoelectricity, internal strain, elastic constant, born effective charges, and dielectric tensor. This phenomenon is related to the Gruneisen parameters, non-linear electric response, anharmonic elastic constants, and third-order phonon-phonon interaction [10]. Observable physical characteristics are listed in order of total energy derivatives. First-order components include the dipole moment, force, and stress tensor. A lot of different things come together to make this happen, including piezoelectricity, internal strain, the elastic constant, born effective charges, and the dielectric tensor. The Gruneisen parameters, nonlinear electric response, anharmonic elastic constants, and third-order phonon-phonon interaction are all linked to this effect. Study that used density functional calculations and thermodynamic models to examine how defects developed in a perovskite-structured oxide in equilibrium with oxygen [11]. Reports that ab initio density functional theory simulations examine pure and nonstoichiometric $BaZrO_3$'s thermodynamic stability, vacancy defect formation energy, and electronic structure. Cation and oxygen vacancies in $BaZrO_3$ insulate and dope holes, affecting its electrical characteristics [12].

# MATERIALS AND METHODES

Design for testing includes IC design methods that include testability into hardware devices. Design for testability of density functional theory is another name for it. Due to the new capabilities, production testing and implementation on the created technology are simpler.

**Hohenberg Kohn Theorem**

The $V_{ext}$ in quantum physics determines the electrical density of the ground state, with the initial state and $V_{ext}$ being proportional. Individuals with real ground state electrical density can recover $V_{ext}$ [13]. Two non-constant body potentials result in different ground states, allowing for the derived Hamiltonians.

$$E_2 = \langle GS2|H_2|GS2\rangle \leq \langle GS1|H_2|GS1\rangle$$

$$\langle GS1|H_2|GS1\rangle = \langle GS1|H_1|GS1\rangle + \langle GS1|H_2 - H_1|GS1\rangle$$

$$= E_1 + \langle GS1|V_2 - V_1|GS1\rangle$$

$$= E_1 + \int (V_2 - V_1)n(r)d3r$$

$$E_2 < E_1 + \int (V_2 - V_1)n(r)d3r$$

Swapping GS1 and GS2 to obtain

$$E_1 < E_2 + \int (V_1 - V_2)n(r)d3r$$

adding above equations they get

$$E_1 + E_2 < E_1 + E_2$$

Hence, there is a one-to-one relationship bettheyen the electronic density at ground state and the exponent $V_{ext}$.

**Local Spin Density Approximation**

The LDA/LSDA correlations and exchange functional are precise in this limit, allowing for an estimate of the Fock exchange integral for homogeneous conduction electrons with a density of n(r). The LDA/LSDA exchange and correlation functional are appropriate for this limit, as the density of a homogeneous gas is (r). Therefore, it is logical to assume that the exchange and correlation functionals are equivalent at some point r [14].

$$Exc = \int \rho(r)\varepsilon x\,(r)Fx\,(\rho(r))d^3r$$

**Exchange and Correlation Approximations**

DFT field exchange and correlation approximations, ranging from local spin density to GW approximation, play a significant role in self-consistency, excluding quantum many body phenomena.

$$Exc = \int \rho(r)vx(r)Fxcd^3r$$

The location and parameters of the enhancement Fxc will change based on the functional type, whether it's LSDA.

**The LSDA+U Method**
Conventional DFT with LSDA cannot accurately predict ground state features of heavily correlated systems, such as semiconducting and half metallic perovskites. GGA typically predicts metallic ground states due to the complex behavior of electrons in transition metals and lanthanides' d and f shells. Approximations within DFT, such as GW, DMFT, and DFT+U, can determine parameters of systems with localized electrons. The LSDA+U strategy, which uses rotation-invariant coordinates and matrix representations of orbital occupancy numbers, is explored using a diagonal representation [15].

$$\Delta E = E_I - E_{dc}$$

The Coulomb matrix consists of elements that can be defined using direct exchange contributions and spin dependent exchange contributions.

$$W_{mm'}\sigma\sigma' = (U_{mm'} - J_{mm'}\delta_{\sigma,\sigma'})$$

The LSDA+U approach considers positive quantities and matrix elements U and J, with the expectation that U will be larger than J. The LSDA technique does not completely eliminate self-interaction, but it can be mitigated by "double counting" correction terms introduced into the energy functioning due to Hubbard U interaction. Other forms of $E_{dc}$, such as the fully localised limit (FLL) and around mean field (AMF), are used in conjunction with LSDA+U functionals. The LDA+DMFT method is another way to count correlation terms. However, the reliability of these methods to a wide range of solid-state materials remains a question.

**Simulation Tools Used in Analysis**

Simcad Pro is a 3D or 2D computer simulation tool that allows users to analyze, optimize, and visualize process flow systems. It also uses WIEN2k, a widely used DFT code, to calculate electronic structures of periodic solids, surfaces, and clusters. The software uses techniques like Full Potential Linearized Augmented Plane Wave (FP-LAPW) for accurate calculations. It can be built using GPL FORTRAN90, C, and MPI compilers. The software also runs sequence flow charts, conducts symmetry operations, and addresses the Eigen value problem for spheres made from muffin tins. The Amsterdam-based ADF Band code, including the periodic DFT code BAND, is used to model periodic systems like crystals, polymers, and slabs. It uses the Linear Combination of Atomic Orbitals (LCAO) and tetrahedron approach for high precision. The EDDMFT simulation software incorporates dynamic mean field theory (DMFT) and DFT, with algorithms for accurate double counting and continuous-time Monte Carlo, one-crossing, and non-crossing approaches for numerical accuracy. GIBBS is a quasi-harmonic model developed by Debye that calculates electronic heat capacity, Gibbs free energy, and Helmholtz free energy by minimizing the non-equilibrium Gibbs function. The Debye gauge's temperature is determined by fitting the Birch Murnagahan equation of state. The Extreme Area Finder for the K Space of Supercells is a new approach for determining ab initio calculations of band energy and de Haas-van Alphen frequencies. It generates a super cell in interpolated k-space, divided into sections for Fermi surface orbits. The Rietveld fitting tool RIETICA can be utilized to determine space group, Wyckoff positions, and lattice parameters using powder diffraction data. It allows tracking of chi-square, Rp, and Rwp patterns, and can be used to fit Rietveld curves to XRD data for LBMO and LMO.

# RESULTS

**Structural Optimization**
The first and most important procedure in computer simulation processing of a material is structural optimization. This is because **"structure dictates properties"** and so obtaining the correct arrangement of electrons wandering in the positive background of the nuclei is critical. The structure can be optimized by including size, form, topology, or both, and so is subject to a wide range of constraint variables. In the case of crystalline materials, geometric optimization is frequently beneficial for determining the shape that minimizes the energy and forces acting on each atom, as well as the strain on a particular system. The disturbance from this geometry may be thought of as the influence of external factors such as pressure and temperature. Geometric optimization may theoretically be achieved by relaxing the structure at different volumes and determining the global minimum energy and right pressure. This can be accomplished by fifing the structure with appropriate equations of state (EoS). There are several models for matching energy or pressure to volume.

**Lattice Dynamics**
In their ground states, atoms in a crystal do not have a fixed position. They are constantly vibrating even at absolute zero, i.e. zero point energy, and respond to external disturbance and thermal energy by increasing their kinetic energy. Atoms display various modes of vibration at higher temperatures, but as temperatures decline, they are constrained to vibrating in fewer and fewer ways. Atoms' displacement from the ground state is smaller than their inter-atomic length.

**Band Structure**
In an isolated atom, the electron energy levels are discretely represented by *1s, 2s, 2p*, … orbitals and electrons occupy these orbitals from the lowest energy levels first ($s < p < d < f$) with Pauli exclusion rule. Each energy level divides into two energy levels of slightly differing energies when two atoms move closer to one another within their interaction distance. In a solid material, the density of atoms (number of atoms per unit cell volume) is in the range of $10^{23}$ atoms/cm$^3$. When a "*N*" number of atoms approach each other to a lattice constant distance, producing a solid, each energy level splits into "*N*" energy levels, forming a continuum. An energy band is a collection of such continuous energy levels. "*2N*" electrons can occupy any energy band.

**Density of States**
We know that the number of energy levels in a band is very large and dependent on the size of a material. Because of this constraint, the number of energy levels or states per unit energy and per unit volume of actual space must be calculated. Analogous to the phonon density of states, the electronic density of states ***D*** is defined as the number of states ***N*** per unit energy ***E*** per unit volume ***V*** of the real space

$$D = \frac{1}{V}\frac{dN}{dE}$$

**Optical Properties**
Optical characteristics are often concerned with a material's capacity to respond to incoming electromagnetic radiation (ER). Materials are characterized as transparent, translucent, or opaque based on this ability. ER causes numerous fascinating features to emerge, depending on whether they are metallic or non-metallic. Recently, scientists discovered novel materials that move when exposed to ER. ER is viewed as a particle rather than a wave, with quantized packets of energy known as photons. The energy E associated with a photon is,

$$E = hV = \frac{hc}{\lambda}$$

On the other hand, the band structure of materials may be used to explain the microscopic treatment of this interaction. Interaction with ER can cause two significant phenomena: (a) electronic polarization and (b) electronic transitions. The electron absorbs the energy given by the input photon, causing ER to slow down, resulting in refraction, which is measured by the refractive index. If the absorbed energy is equal to or larger than the band gap, electronic transitions may occur, and the difference in energy between the unexcited and excited states is,

$$\Delta E = h\nu$$

The excited electron emits a photon before returning to its unexcited state, resulting in another phenomena known as luminescence. The frequency of electromagnetic radiation released by a substance determines its color.

Figure 1 shows the absorption spectra of barium zirconate perovskites calculated by DFT. Figure shows the pure structure of barium zirconate supercell as reference structure along with supercells dopped with nitrogen atoms just by replacing the oxygen atoms from the lattice sites. The graph shows significant change in the absorption spectra by incorporating various (2, 4, 6 and 8) numbers of nitrogen atoms in the barium zirconate supercell.

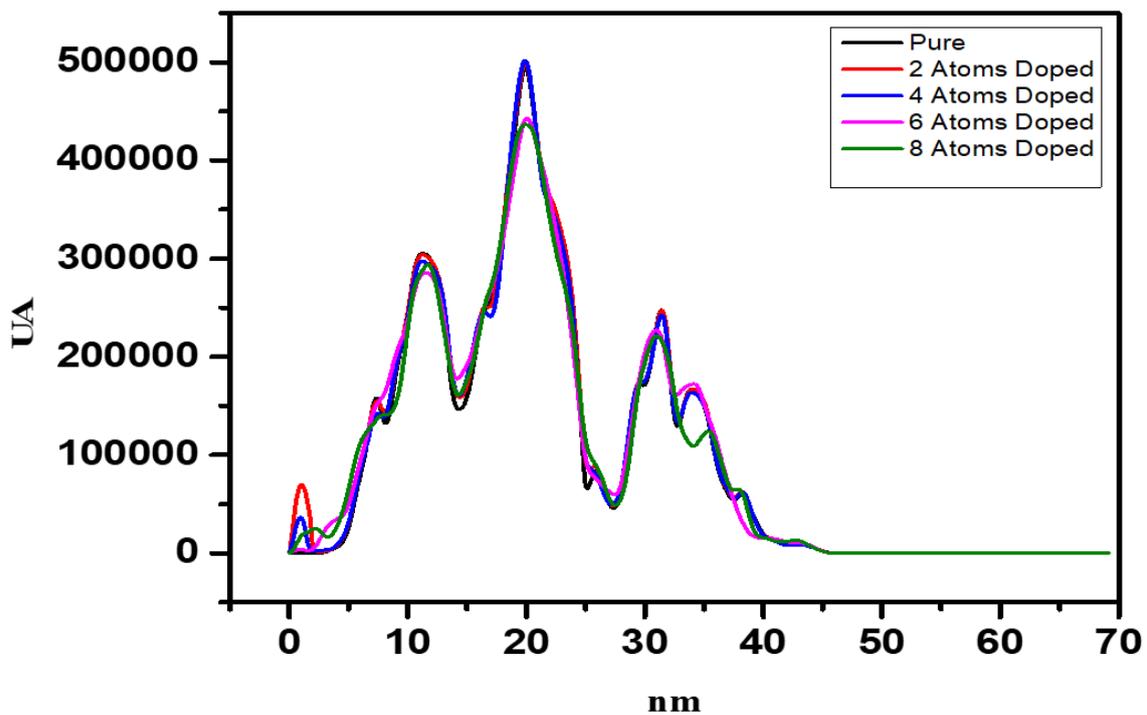

**Figure 1** DFT study of the effect of substituents on the absorption spectra

The PDOS graphs of barium zirconate perovskites calculated by DFT are shown in Figure 2. Figure 2 depicts the pure structure of barium zirconate as well as graphs of nitrogen atom doping by substituting oxygen atoms in the lattice locations. The image depicts a dramatic variation in the PDOS graphs caused by integrating different quantities of nitrogen atoms in the barium zirconate supercell.

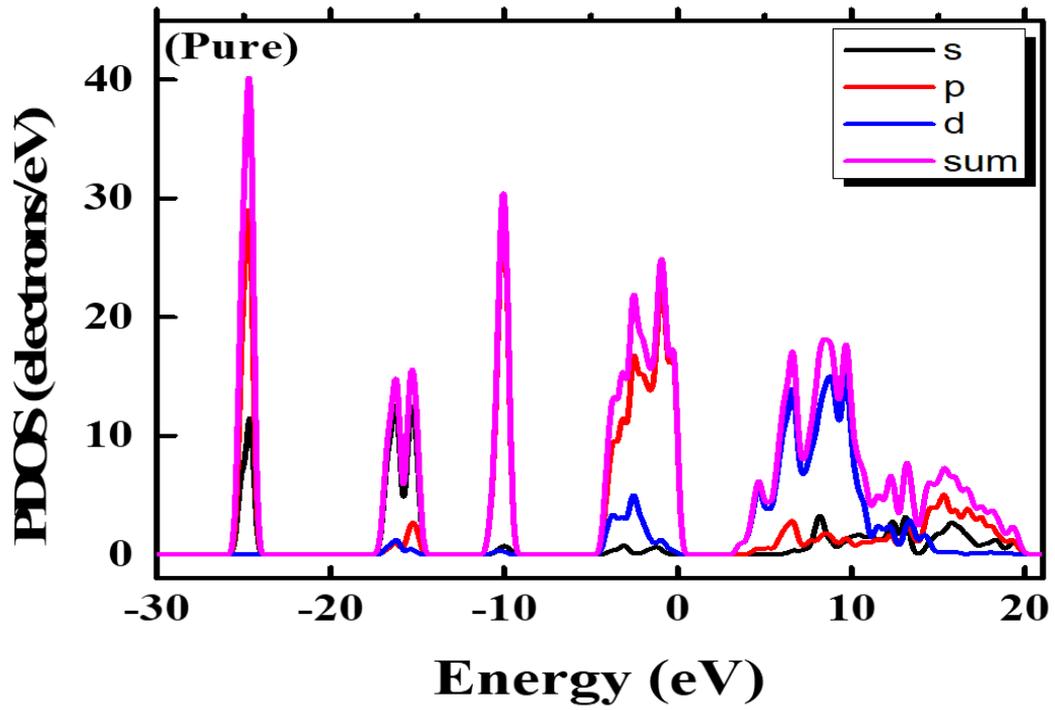
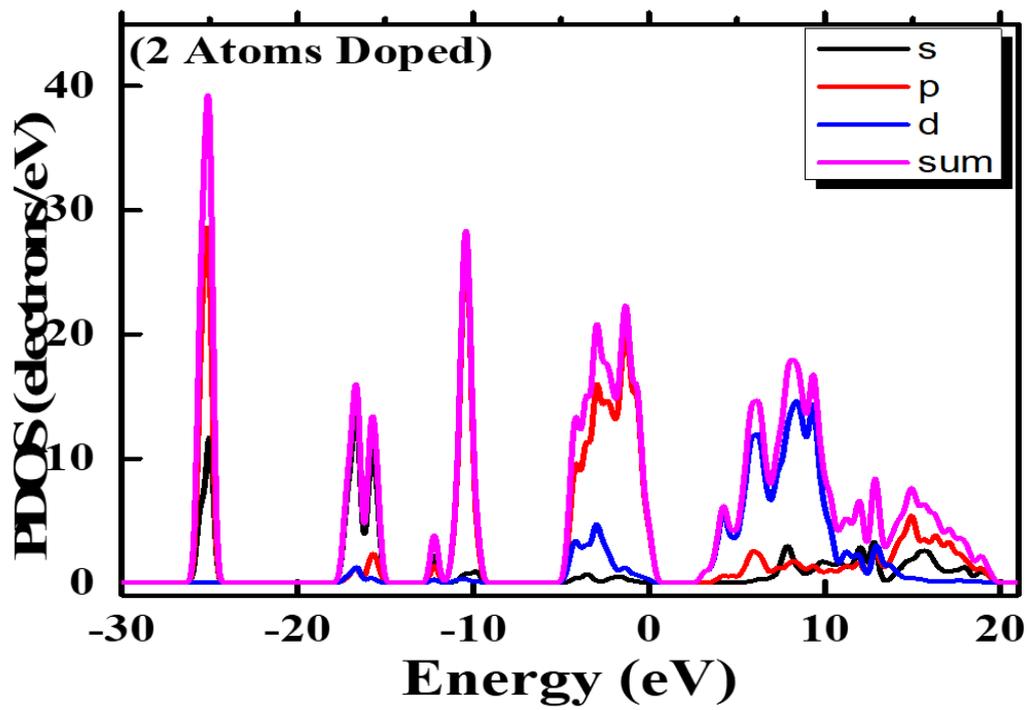

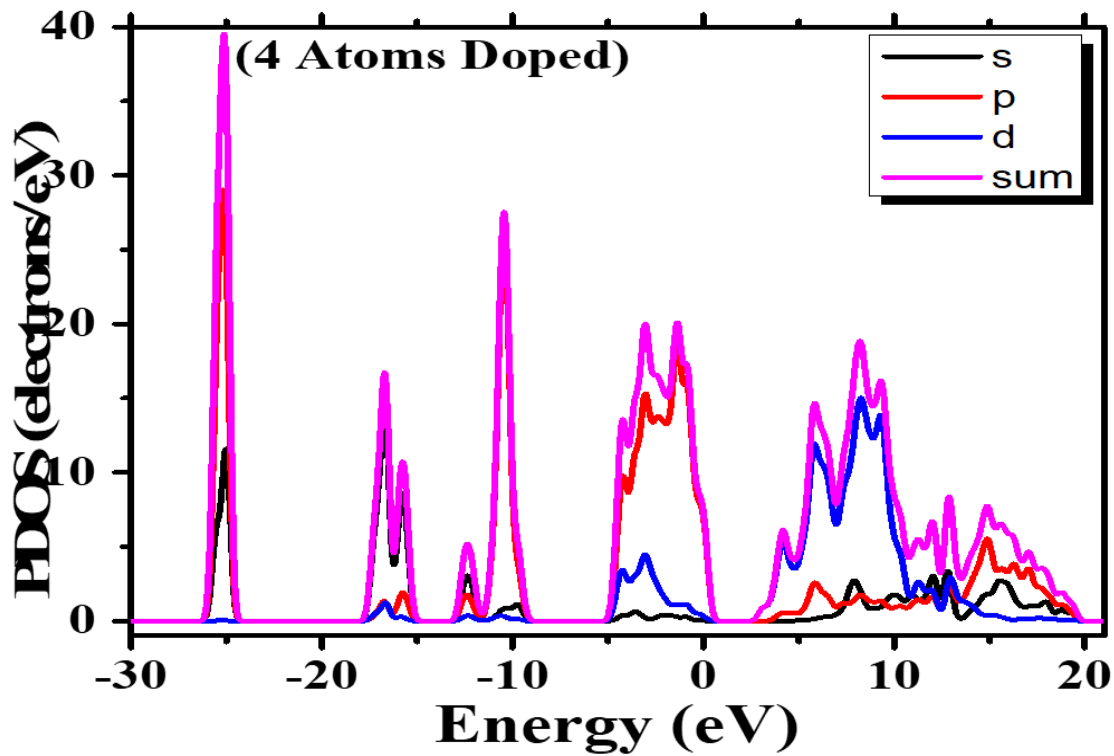

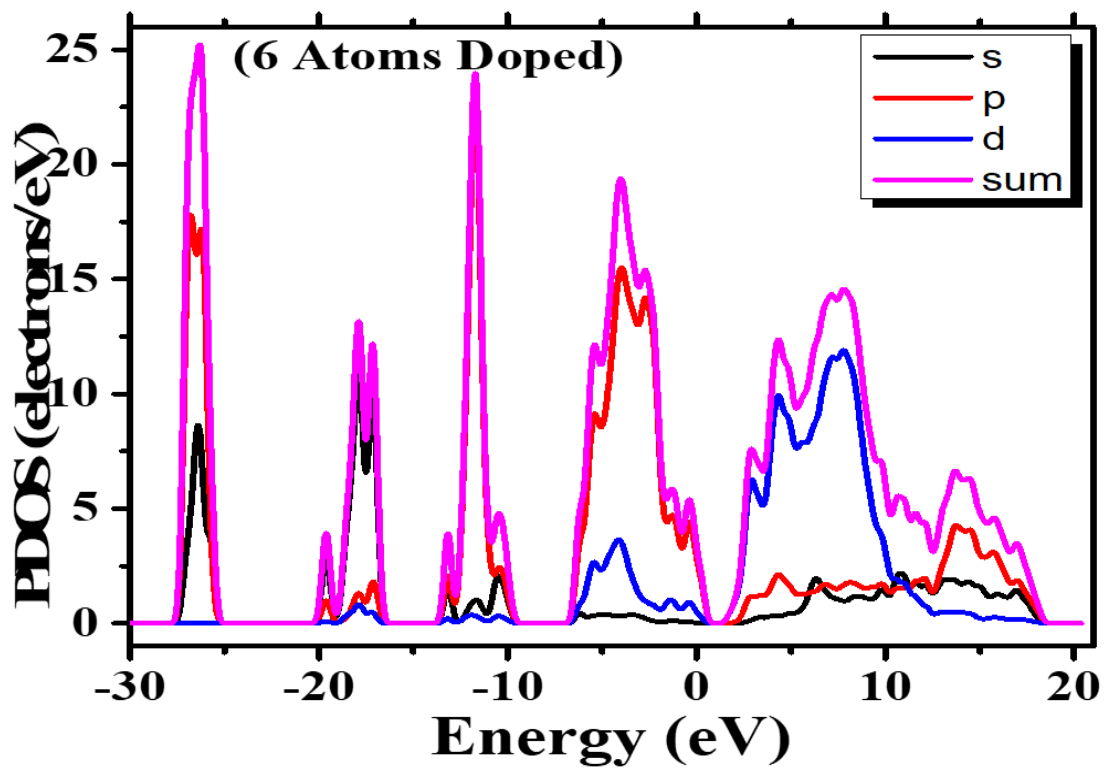

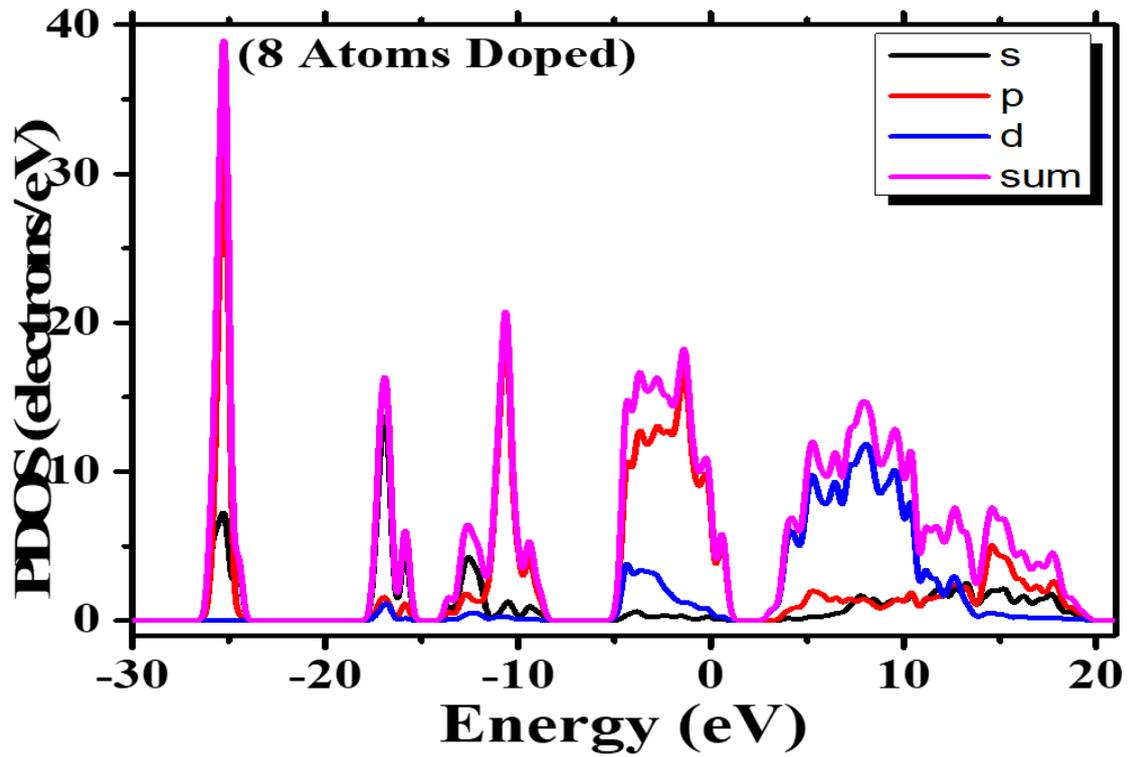

**Figure 2** PDOS graphs of the Barium Zirconate supercell dopped with Nitrogen atoms (2, 4,6,8 along with pure)

The predicted phonon band structure of barium zirconate perovskites is shown in Figure 3. This picture depicts the pure structure of barium zirconate, followed by the substitution of nitrogen atoms for oxygen atoms in the lattice positions. The image depicts a dramatic variation in the band structure caused by inserting varying quantities of nitrogen atoms in a barium zirconate supercell.

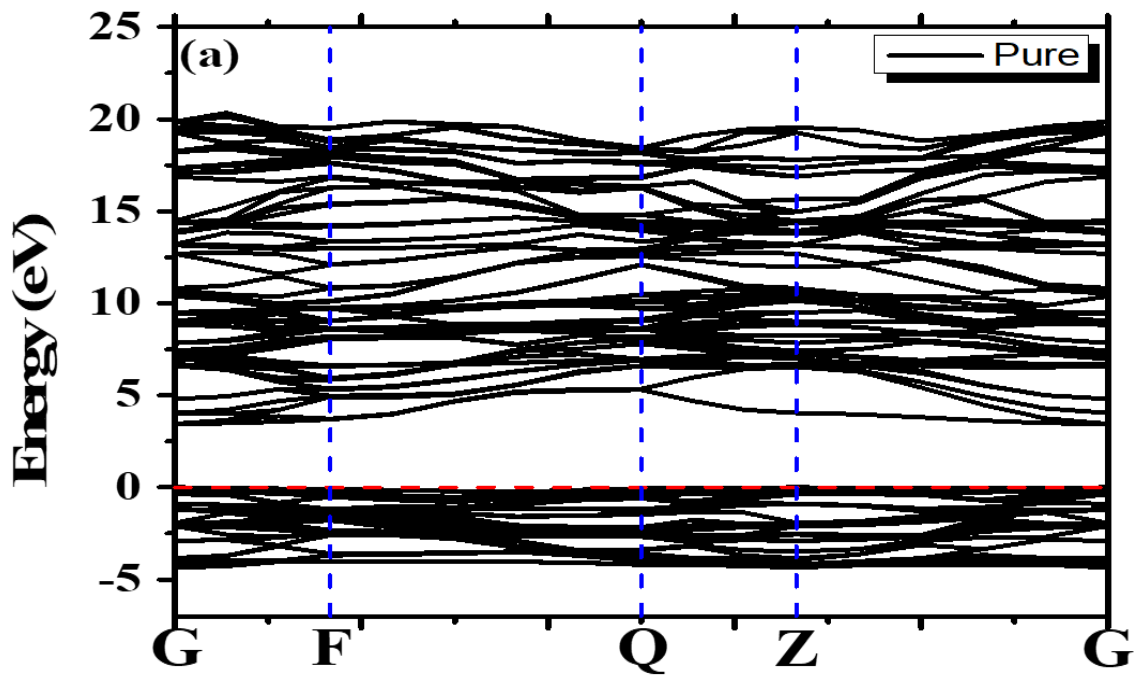
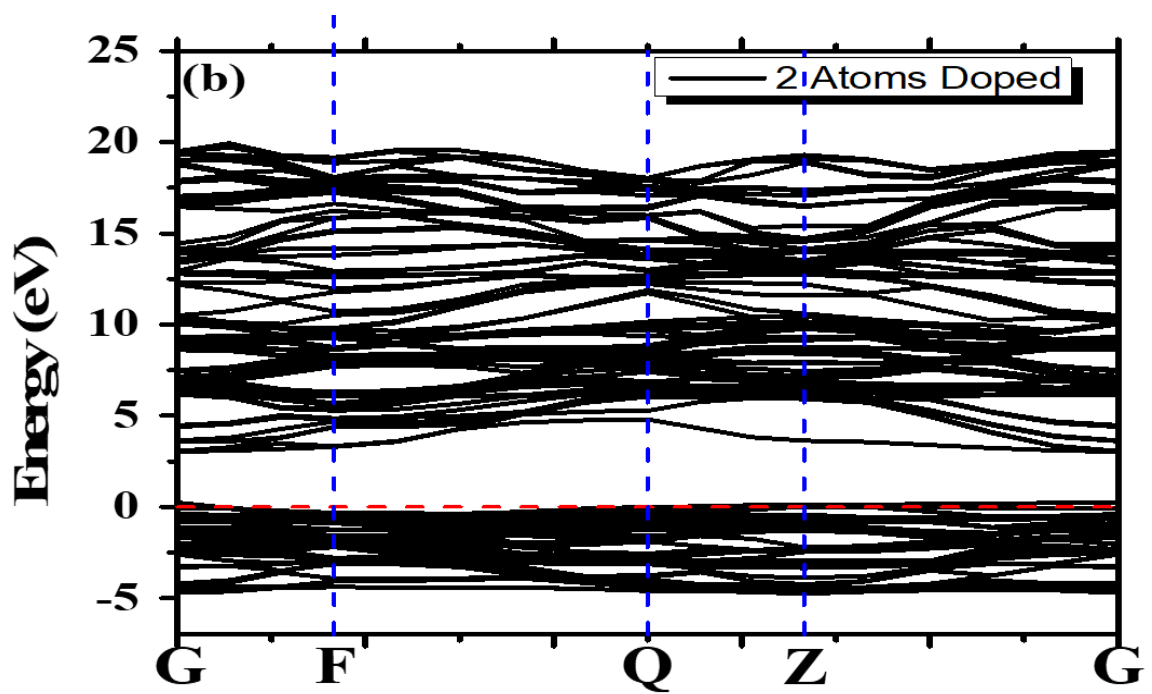

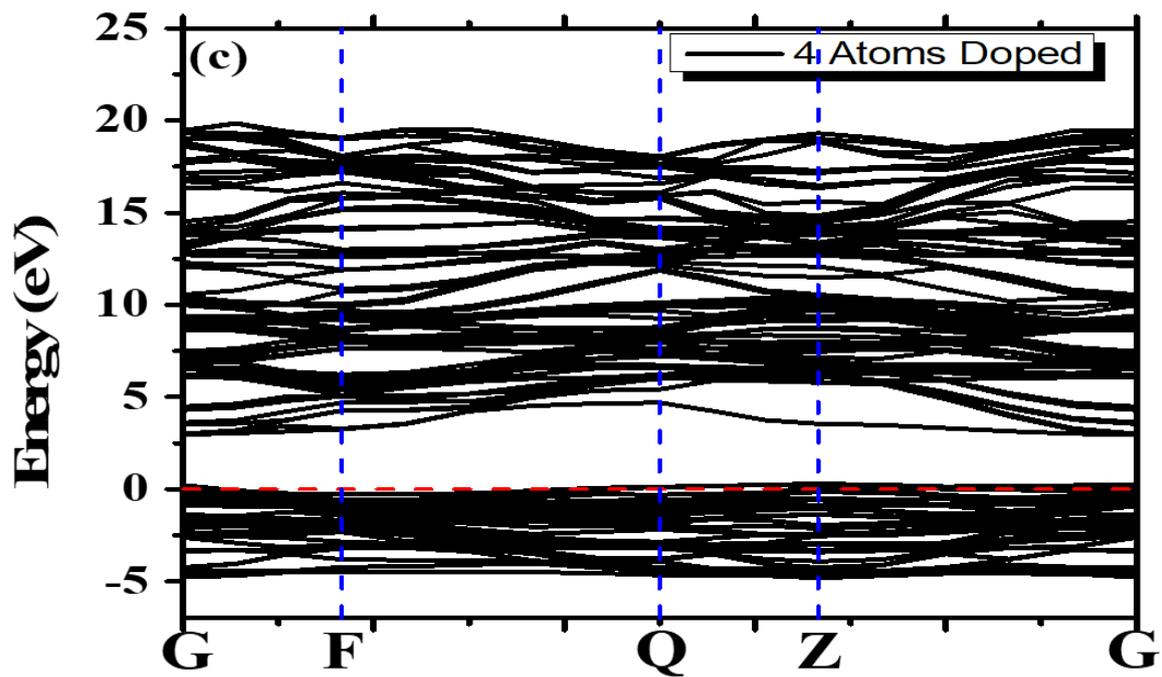

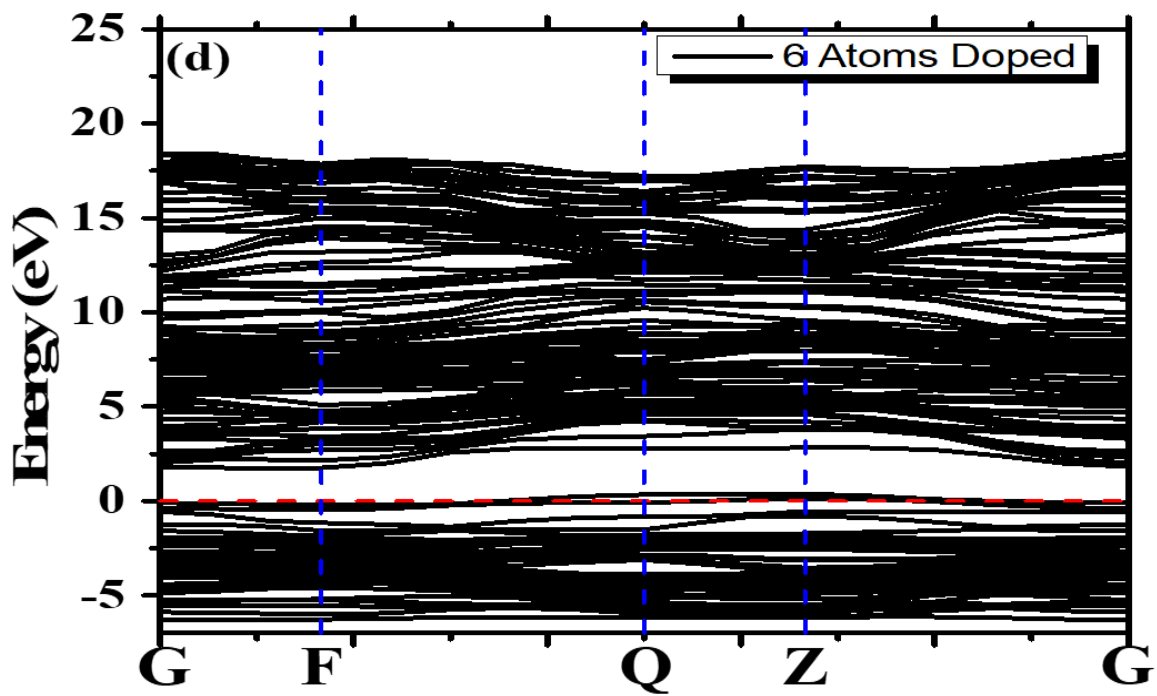

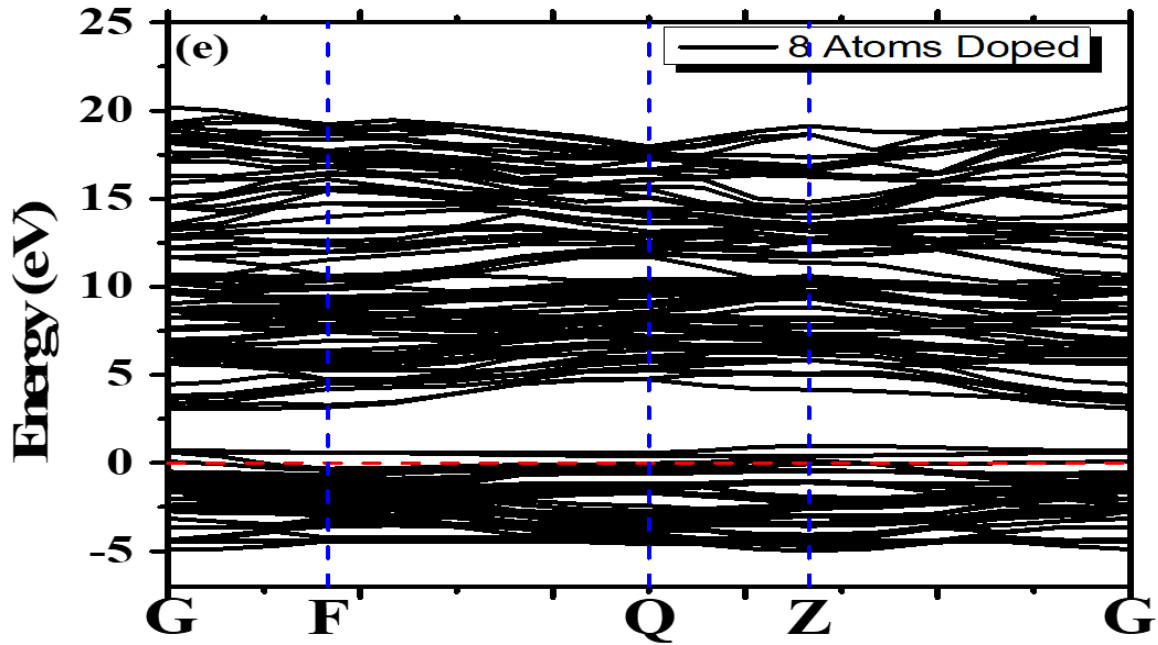

**Figure 3** Band structures of the Barium Zirconate supercell dopped with Nitrogen atoms (2, 4,6,8 along with pure)

## DISCUSSIONS

Using density functional theory (DFT), we analyse how nitrogen doping affects the structural, electrical, and optical aspects of barium zirconate (BZ). The findings show a considerable decrease in bandgap energy, better visible light absorption, and the introduction of nitrogen induced states in the bandgap area. These findings indicate that nitrogen doping can improve the photocatalytic activity of BZ, making it a suitable material for water splitting and pollutant degradation. Previous research have described the effective doping of nitrogen into BZ using experimental and theoretical approaches, which is consistent with our results. For example, created nitrogen doped BZ nanoparticles and discovered increased photocatalytic activity for water splitting. Similarly, conducted DFT simulations to study the impact of nitrogen doping on the electrical structure of BZ and discovered a decrease in bandgap energy [16]. These experiments show how nitrogen doping can improve the photocatalytic capabilities of BZ. Our investigation into the structural, electrical, and optical properties of nitrogen-doped barium zirconate (BZ) was carried out with the assistance of density functional theory (DFT), which was utilized for this study. The findings of our research provide us with important insights into the influence that nitrogen doping has on the behavior of the material, which is of tremendous value for the future uses of the material. BZ that has been doped with nitrogen has had its crystal structure optimized, and the results demonstrate that the lattice parameters have expanded slightly. This expansion is a natural consequence of the fact that nitrogen possesses a greater ionic radius than oxygen. Through careful examination, it is possible to see that this expansion has led to a decrease in the density of the material, which may have an effect on the material's thermal and mechanical properties. An investigation into the structure of electronic bands and density of states (DOS) calculations revealed that nitrogen doping significantly reduced the bandgap's energy. This indicates that there has been an improvement in both optical absorption and conductivity. The incorporation of nitrogen-induced states into the bandgap area paves the way for the possibility of employing these states in optoelectronic devices and photocatalysis, both of which are very interesting prospects.

Other investigations, however, have found that nitrogen doping has varying impact on the characteristics of BZ. For example, discovered that nitrogen doping enhanced the bandgap energy of BZ, which they ascribed to the creation of nitrogen related defects [17]. This disparity emphasises the complexities of nitrogen doping in BZ and the necessity for more research to properly understand its effects. Comparison to other research work our findings are comparable with those of, who found improved photocatalytic activity in nitrogen-doped BZ nanoparticles [16]. Our findings are consistent with the theoretical analysis of, who anticipated a decrease in the bandgap energy of BZ with nitrogen doping. However, we observed an increase in the bandgap energy of BZ after nitrogen doping, indicating the need for further research to resolve this discrepancy. In general, our study adds to what is already known by giving a full picture of how adding nitrogen changes the electrical, optical, and structural features of BZ using density functional theory (DFT) technology. The findings could significantly impact both optoelectronic and photocatalytic applications, with major implications for the design of BZ-based materials. An excited electron emits a photon before returning to its unexcited state, triggering luminescence. A substance determines its color by the frequency at which it emits electromagnetic radiation. Figure 1 displays the DFT measurements of the absorption spectra of barium zirconate perovskites. The figure serves as a reference, showing both a pure barium zirconate supercell and supercells doped with nitrogen atoms by modifying oxygen atoms at lattice locations. The graph clearly demonstrates that adding nitrogen atoms 2, 4, 6, and 8 to the barium zirconate supercell composition significantly transforms the absorption spectrum. Figure 2 presents the PDOS graphs of barium zirconate perovskites, derived from the density functional theory. Figure 2 depicts the structure of pure barium zirconate and graphs the doping effects of nitrogen atoms on oxygen lattice positions. When adding a variable quantity of nitrogen atoms to barium zirconate supercells, the PDOS graphs undergo a significant shift. Figure 3 depicts the expected phonon band structure in barium zirconate perovskites. Here, we showcase the pure structure of barium zirconate, subsequently leading to the substitution of oxygen atoms for nitrogen atoms throughout the lattice. The graphic shows how adding nitrogen atoms to a barium zirconate supercell alters its band structure.

## CONCLUSION

DFT computations were used to determine the development of protonic, nitrogen ionic, and electronic conductivities in Barium Zirconate. To begin, nitrogen doping effectively lowered the energy barriers of proton and oxygen ion migration. The electronic band structure suggests that nitrogen doping resulted in poor electronic conductivity. Second, as the nitrogen level of the supercell increases, so does the conductivity of the Barium Zirconate. Finally, the tendency of the measured total conductivity is compatible with the DFT calculation results. Furthermore, Density Functional Theory was used to dope $BaZrO_3$ in order to explore the local configurations of the octahedral sites in Pm3m cubic frameworks. Nitrogen ion substitution in the $BaZrO_3$ supercell was investigated, including the existence of an oxygen vacancy nearby. Although the structural symmetry of undoped $BaZrO_3$ was not altered by nitrogen doping, the presence of nitrogen caused numerous changes in the oxygen sites surrounding it based on nitrogen's local geometrical arrangement in the host matrix. The protonated pieces in this example had differing stabilization energies. Only when two nitrogen atoms were nearby were the relative energy differences between the individual proton stable sites in line with the order of magnitude of the observed proton-hopping activation energies. The distribution of such energy differences indicated a three-grouping of oxygen atoms, each having different structural features that could not be derived from their topologies. Using energy-difference distributions, the occurrence of proton traps was also investigated.